\documentclass[preprint,floats,aps,epsfig,nofootinbib,amssymb]{revtex4-1}
\usepackage{mathrsfs}

\usepackage{slashed}
\usepackage{graphicx,color}
\usepackage{dcolumn}
\usepackage{bm}
\usepackage{subfig}
\usepackage{graphicx}
\usepackage{amssymb}
\usepackage{stackrel}
\usepackage{pgfplots}
\usetikzlibrary{positioning}
\usetikzlibrary{snakes}

\begin{document}

\title{Pseudo-Dirac Dark Matter in XENON1T}

\author{Wei Chao$^1$}
\email{chaowei@bnu.edu.cn}
\author{Yu Gao$^2$}
\email{gaoyu@ihep.ac.cn}
\author{Mingjie Jin$^1$}
\email{jinmj@bnu.edu.cn}
\affiliation{$^1$Center for Advanced Quantum Studies, Department of Physics, Beijing Normal University, Beijing, 100875, China \\
$^2$Key Laboratory of Particle Astrophysics, Institute of High Energy Physics,
Chinese Academy of Sciences, Beijing 100049, China
}
\vspace{3cm}

\begin{abstract}
The XENON1T dark matter experiment recently reported  0.65 ton-year exposure measurement on electron recoils , which shows an excess in $2\sim 3$ KeV  recoils above the detector background. In this paper we present a Pseudo-Dirac dark matter scenario to explain the excess via inelastic dark matter-electron scattering.  With a KeV scale mass splitting between the two components of the Pseudo-Dirac dark matter, the slightly excited component can down-scatter on electrons. The desired dark matter masses are about 10 GeV  with a 4 KeV mass-splitting and unity coupling to electrons, which generate the observed XENON1T recoil events, give the appropriate  dark matter relic abundance and satisfy collider search limits.
\end{abstract}

\maketitle
\section{Introduction}

Astronomical observations have confirmed the existence of dark matter (DM), which cannot be explained in the framework of the minimal Standard Model (SM).  For the past decades, many dark matter models have been created, of which some typical DM scenarios, such as Weakly interacting massive particles (WIMP), axion, sub-GeV DM and primordial   black hole, come to the foreground.  To probe the nature of the DM, one can detect the  recoil energy of nuclei or electron arising from the collisions of nuclei with WIMP in underground laboratories, detect the secondary cosmic rays from the annihilation or decay of the DM, and produce the DM directly at the LHC.  Benefiting from technological advances, great strides have been made in improving the detection precision and efficiency.  We are in the age of experimental data explosion.  

Recently the XENON1T experiment~\cite{Aprile:2020tmw} reported results from searches for DM with low-energy electronic recoil data acquired from February 2017 to February 2018.  There are $285$ events observed  in the electron recoil energy between 1 keV and 7 keV, against an expected background of  $232\pm15$ events, corresponding to a $3.5\sigma$ Poisson fluctuation.  The excess may be explained by either the solar axion  or the neutrino magnetic moment~\cite{Aprile:2020tmw}, but both are disfavored by existing astrophysical constraints~\cite{Giannotti:2017hny,Corsico:2014mpa}. A third explanation is the beta decay of tritium, which cannot be confirmed yet due to the lack knowledge of the tritium property.  Some other new physics explanations have proposed to explain the XENON1T excess, which can be classified into the following five categories: (a) boosted DM-electron scattering~\cite{Fornal:2020npv,Chen:2020gcl,Du:2020ybt,Su:2020zny,Primulando:2020rdk,Cao:2020bwd,Dey:2020sai,Zu:2020idx}; (b) exotic neutrino interactions~\cite{Boehm:2020ltd,AristizabalSierra:2020edu,Bally:2020yid,Lindner:2020kko}; (c) absorption of  bosons such as axions, dark photons~\cite{Buch:2020mrg,DiLuzio:2020jjp,Alonso-Alvarez:2020cdv,Choi:2020udy,Nakayama:2020ikz,An:2020bxd,Gao:2020wer,Kannike:2020agf,Takahashi:2020bpq};   (d ) ``fake" photon signals  from annihilation or decay of DM~\cite{Paz:2020pbc,Bell:2020bes}; (e) Inelastic DM-electron scattering~\cite{Harigaya:2020ckz,Lee:2020wmh,Baryakhtar:2020rwy,Bramante:2020zos}.

In this paper we propose an explanation to the XENON1T excess with a Pseudo-Dirac DM $\chi$, where the Pseudo-Dirac fermion  consist of two Majorana fermion components, $\eta_{a,b}$, with a tiny mass splitting. The concept of Pseudo-Dirac fermion is introduced to address  the problem of  tiny Majorana masses of active neutrinos via the inverse seesaw mechanism~\cite{Mohapatra:1986bd}. Here we take the Pseudo-Dirac fermion, which can be stabilized by a discrete symmetry or a global U(1) symmetry, as  two-component DMs that interact with the right-handed electron via a charged scalar singlet. We focus on constraints on the model from the observed relic abundance and the direct detection signals, while eluding indirect detection as previously studied in Ref.~\cite{Chao:2019xji}.  We find that the down-scattering process $\eta_b e\to \eta_a e$, where $m_{\eta_b} > m_{\eta_a}$ can address the excess in the low energy electron recoil data.   The desired DM mass and mass splitting is about $10$ GeV and $4$ keV, respectively.  Constraints on the model from the observed relic density is also studied. 

The paper is organized as follows: In section II we present the model in detail. In section III we calculate the relic density of the Pseudo-Dirac DM. Section IV is devoted to the study of DM-electron scattering. The last part is concluding remarks.

\section{Pseudo-Dirac Dark Matter}

Pseudo-Dirac fermion are introduced in explaining the  neutrino tiny Majorana mass via the inverse seesaw mechanism~\cite{Cai:2014hka}.  In this section we present a pseudo-Dirac DM model, which extends the SM with a Pseudo-Dirac fermion $\chi$ and a charged scalar singlet $\varphi$. $\chi$ is stabilized by a $Z_2$ symmetry, and is thus a cold DM candidate.  Assuming $\chi $ mainly couple to right-handed electron,  the Lagrangian can be written as 
\begin{eqnarray}
-{\cal L}  \sim \overline{\chi_L^{}} M \chi_R^{} + {1\over 2} \mu \overline{\chi_L^{} }  \chi_L^C + \zeta \overline{\chi_L^{} } \varphi^+ E_R^{} + {\rm h.c.} 
\end{eqnarray}
where $M$ is the Dirac mass, $\mu$ is a tiny Majorana mass parameter, $\zeta$ is the Yukawa coupling between $\chi$,  a charged scalar $\varphi^+$ and the  right-handed electron $E_R$.  We will not stress on the naturalness problem of a small Majorana mass term in this paper, which can be addressed by 't Hoofts' naturalness principle~\cite{thooft}.  In the basis $(\chi_L^{}, ~\chi_R^C)$, the DM mass term can be written as
\begin{eqnarray}
{1\over 2} \overline{\left(\matrix{\chi_L^{} & \chi_R^C}\right)}\left(\matrix{ \mu &M \cr M & 0}\right) \left( \matrix{ \chi_L^C \cr \chi_R^{} } \right) + {\rm h.c.} \; . 
\end{eqnarray}
We define the symmetric mass matrix as ${\cal M}$, it can be diagonalized by a $2\times 2$ unitary transformation 
\begin{equation}
{\cal U}=\left(\begin{array}{cc}
c & -s \\
s & c
\end{array}\right) ,
\end{equation}
i.e., ${\cal U}^\dagger {\cal M} {\cal U}^* = {\rm diag} \{m_1^{} , m_2^{} \}$ and we denote $c=\cos\theta$ and $s=\sin\theta$.   In the limit $\mu \ll M$, the mass eigenvalues and the mixing angle can be written as 
\begin{eqnarray}
\tan \theta = 1-\alpha \; , \hspace{1cm} m_1 \approx M-\mu \; , \hspace{1cm} m_2 \approx M+\mu \; ,  
\end{eqnarray}
where $\alpha ={\mu/2M}$. We define the mass eigenstates as $\eta^a$ and $\eta^b$, where $\eta^\kappa =\eta_L^{\kappa} + \eta_L^{\kappa C}$ $(\kappa=a,b)$, then the interaction eigenstates and mass eigenstates are related by
\begin{eqnarray}
\chi_L^{} = c \eta^a_L -s \eta^b_L \; , \hspace{1cm} \chi_R^{} =s \eta^{aC}_L+ c \eta^{bC}_L \; ,
\end{eqnarray}
and the portal interaction takes the form 
\begin{eqnarray}
\zeta (c \overline{\eta^a_L} -s \overline{\eta^b_L} ) \varphi^+ E_R^{} + h.c. \; ,
\end{eqnarray}
which means that we have two components Majorana dark matter with tiny mass splitting in this model, which forms a Pseudo-Dirac fermion. 

\section{Relic abundance}
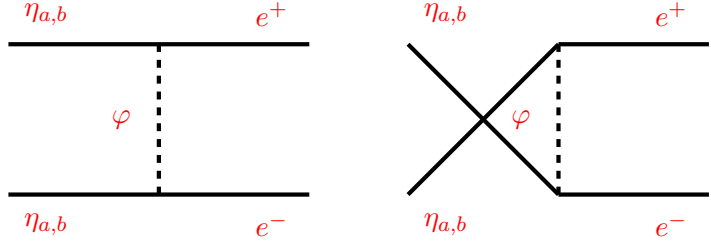
\begin{figure}
\begin{center}
\begin{tikzpicture}
\draw[-,ultra thick] (-1,0)--(3,0);
\draw [-,dashed, ultra thick] (1,0) -- (1,2);
\draw[-,ultra thick] (-1,2)--(3,2);
\node[red, thick] at (-0.5,-0.4) {$\eta_{a,b}$};
\node[red, thick] at (2.5,-0.4) {$e^-$};
\node[red, thick] at (0.5,1.0) {$\varphi$};
\node[red, thick] at (-0.5,2.4) {$\eta_{a,b}$};
\node[red, thick] at (2.5,2.4) {$e^+$};
\end{tikzpicture}
\hspace{1cm}
\begin{tikzpicture}
\draw[-,ultra thick] (-1,2)--(1,0);
\draw[-,ultra thick] (1,0)--(3,0);
\draw [-,dashed, ultra thick] (1,0) -- (1,2);
\draw[-,ultra thick] (1,2)--(3,2);
\draw[-,ultra thick] (-1,0)--(1,2);
\node[red, thick] at (-0.5,-0.4) {$\eta_{a,b}$};
\node[red, thick] at (2.5,-0.4) {$e^-$};
\node[red, thick] at (0.5,1.0) {$\varphi$};
\node[red, thick] at (-0.5,2.4) {$\eta_{a,b}$};
\node[red, thick] at (2.5,2.4) {$e^+$};
\end{tikzpicture}
    \caption{Feynman diagrams for the annihilation of DM}\label{feynman}
\end{center}
\end{figure}
As illustrated in section II, both $\eta_a$ and $\eta_b$ couple to the visible sector and become dark matter candidates.  With a sizable coupling $\zeta$, $\eta^{a,b}$ are in thermal equilibrium with the thermal bath in the early universe, then freeze-out as their interaction rates drop below the Hubble rate.  The number densities  $n_{a,b}$ evolve according to the Boltzmann equations,
\begin{eqnarray}
\dot{n}_a + 3 Hn_a &=& -\gamma(\bar \eta_a  \eta_a \to e^+e^-) \left(  {n_a^2 \over n_{a, eq}^{2}} -1\right) -\gamma(\bar \eta_b  \eta_a \to e^+e^-) \left(  {n_a n_b \over n_{a, eq}^{}n_{b, eq}} -1\right)\\
\dot{n}_b+3 H n_b &=&-\gamma(\bar \eta_b  \eta_b \to e^+e^-) \left(  {n_b^2 \over n_{b, eq}^{2}} -1\right) -\gamma(\bar \eta_a  \eta_b \to e^+e^-) \left(  {n_a n_b \over n_{a, eq}^{}n_{b,eq}} -1\right)
\end{eqnarray}  
with
\begin{eqnarray}
\gamma(ab\to cd ) ={T\over 512 \pi^6} \int |{\cal M}|^2 p_{ab} p_{cd} {1\over \sqrt{s}} K_1\left( {\sqrt{s} \over T}\right) d\Omega ds 
\end{eqnarray}
where $K_1 (x)$ is the Bessel function of the second kind of oder 1, $|{\cal M}|^2$ is the squared
amplitude of the process $ab\to cd $, $p_{ij}$ is the momentum of particle $i$ and $j$ in the center-of-mass frame,
\begin{eqnarray}
p_{ij} ={1\over 2 \sqrt{s}} \sqrt{s-(m_i+m_j)^2} \sqrt{s-(m_i-m_j)^2} \; .
\end{eqnarray}
Feynman diagrams relevant for the annihilation of DM are given in the Fig.~\ref{feynman}.  One has$n_a \approx n_b$ for  $\theta \approx \pi/4$.  Relic abundance is obtained by solving eqs.(6) and (7) numerically.

\begin{figure}
\centering
\includegraphics[width=0.6\textwidth]{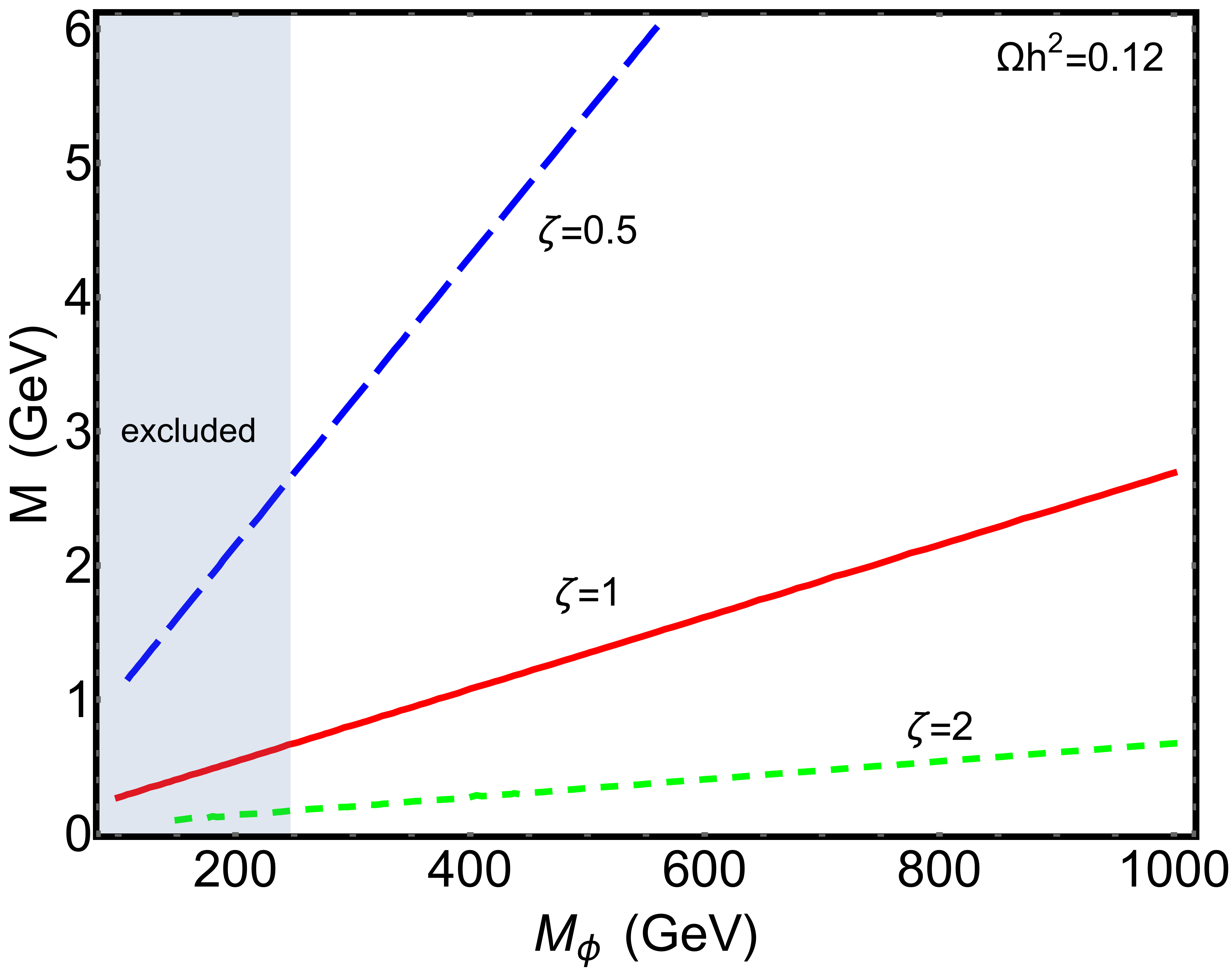}
\caption{ Contours of the DM relic density in the $M_\phi-M$ plane, where the solid, dashed and dotted lines correspond to $\zeta=1,~0.5$ and $2$ respectively. The shaded region is excluded by LHC search~\cite{Sirunyan:2018nwe}. 
} \label{relicD}
\end{figure}

Fig.~\ref{relicD} shows in the $M_\varphi-M$ plane contours for the observed relic abundance $\Omega_Mh^2=0.12$,  with the solid, dashed and dotted lines correspond to $\zeta=1,~0.5$ and $2$ respectively. The shaded regime is ruled out by the latest CMS search on $E_R$'s heavy partners\cite{Sirunyan:2018nwe} .  
 With $\theta$ close to $\pi/4$ in this model,  the interaction strength of $\eta_a$  and $\eta_b$ are almost the same, and the DM relic density becomes insensitive to the tiny mass splitting.  DM with GeV-scale masses can obtain correct abundance with $\zeta \sim {\cal O}(0.1-1) $. 

\section{Pseudo-Dirac DM -electron scattering}

\begin{figure}
\centering
\includegraphics[width=0.6\textwidth]{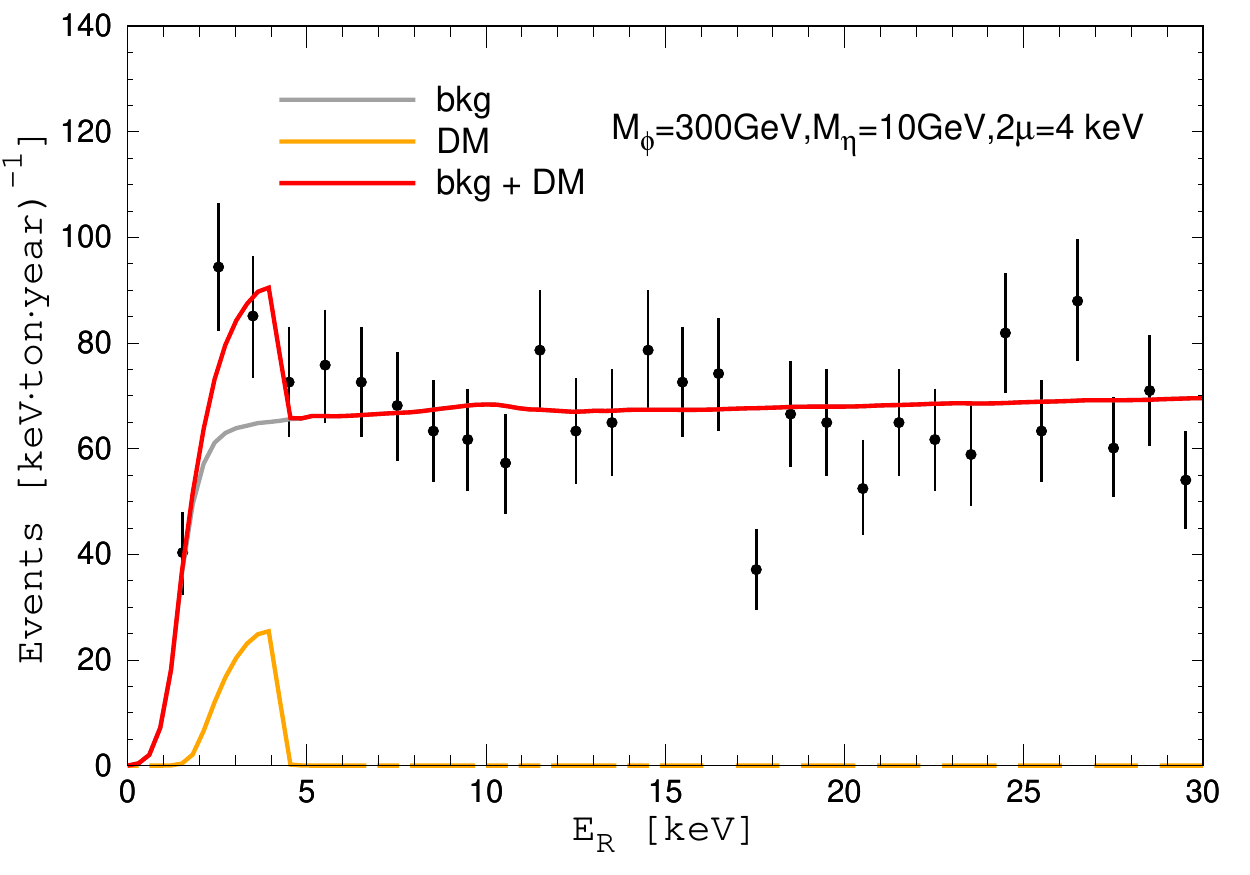}
\caption{Electron recoil spectrum from XENON1T. The orange line represents the contribution from DM-electron scattering with a benchmark $m_\eta$=10 GeV, $m_\varphi$=300 GeV and $\mu$=2 keV. The red line shows the total electron energy distribution. Here the Yukawa couplings $\zeta$ is unity.} 
\label{sigbkg}
\end{figure}

The matrix element between DM and  electron mediated by $\varphi$ takes the form
\begin{eqnarray}
i{\cal M} &=& (-i\zeta)^2  (c \overline{\eta_a} -s\overline{\eta_b}) P_R^{}  e^{} { i\over p^2 -M_\varphi^2 } \bar e P_L (c \eta_a -s\eta_b) \nonumber \\
&\approx&  {- i \zeta^2 \over M_\varphi^2 -(M+M_e)^2} {cs \over 8} \left(\bar \eta_a \gamma_\mu \eta_b + \bar \eta_b^{} \gamma_\mu \eta_a^{} \right)   \bar e \gamma^\mu e +\dots
\end{eqnarray}
where the propagator is expanded in the low momentum limit and only the leading order term are kept, Fierz transformation is performed in the second line and axial-vector currents are neglected. Due to the Majorana nature, $\bar \eta_a \gamma_\mu \eta_a =\bar \eta_b \gamma_\mu \eta_b =0$.   The  free electron scattering cross section at the momentum transfer $q=a_0^{-1} =\alpha m_e$, where $a_0$ is the Bohr radius and $\alpha$ is the fine structure constant,  can then be written as
\begin{eqnarray}
\sigma_0 \approx {\zeta^4 \mu^2 \over 64 \pi M_\varphi^4}
\end{eqnarray}
where $\mu$ is the DM-electron reduced mass and we have taken $cs\approx {1/2}$ in the calculation. The differential rate due to scattering between $\eta$ and the electron is
\begin{eqnarray}
{d \sigma v \over d E_R} =  { \sigma_0 \over 2 m_e } \int \frac{d v_\eta f(v_\eta)}{v_\eta}  \int_{q^-}^{q^+} a_0^2 q dq K(E_R, q) |F(q)|^2
\end{eqnarray}
where $K(E_R, q)$ is the atomic excitation factor~\cite{Roberts:2016xfw,Roberts:2019chv}, $F(q)$ is the dark matter form factor, and $q$ is the transferred momentum. For $E_R\sim 2~{\rm keV}$, one has $K\sim0.1$. We take $F(q)=1$ in our calculation and the integration limits $q_\pm$ are derived by solving the following equation,
\begin{eqnarray}
 {q^2  } -{2pq\cos\Theta } = 2 m_1 (2 \mu- E_R ) + p^2 \left({m_1 \over m_2 } -1\right)   
\end{eqnarray}
where $\Theta$ is the angle between the momentum of the incoming DM and transferred momentum, $p$ is the momentum of the incoming DM.  By assuming $2\eta> E_R +p^2(m_1^{-1} -m_2^{-1})/2$, the limits of the integration $q^\pm$ can be written as
\begin{eqnarray}
q^{\pm} =\pm m_2 v + \sqrt{ m_1^{} m_2^{}  v^2 +2m_1 (2\mu -E_R )} 
\end{eqnarray}
which will be used in our simulations.   For the other case, $q^\pm$ are
\begin{eqnarray}
q^{\pm} = m_2 v \pm \sqrt{ m_1^{} m_2^{}  v^2 +2m_1 (2\mu -E_R )} 
\end{eqnarray}
which  recover the values for elastic scattering in the limit $\eta\to0$.

The total number of events take the form
\begin{eqnarray}
n_e= M_{T} T  N_T n_{\eta_b } \int dE_R^{}\int dE' \varepsilon(E^\prime)  {\cal G}(E_R,E')^{}{\cal F}(E')\cdot \left.{d \sigma v \over d E' }\right.,
\end{eqnarray}
where ${\cal F}=\sum_{i}\theta(E_R-B_i)$ is a good approximation of corrections due to Xe atomic bindings~\cite{Chen:2016eab} in which $B_i$ is the $i$th electron's binding energy. $M_{T}$ is the target mass, $T$ is the exposure time and $M_T T =0.65~{\rm ton\cdot year}$, $N_T\approx 4.2\times 10^{27}~{\rm ton}^{-1}$ being the number of atoms in the target, $n_{\eta_b}\approx 0.2 ~{\rm GeV\cdot cm^{-3}}$ being the number density of the $\eta_b$, ${\cal G}$ is a $\sim 2$ KeV Gaussian smearing on recoil energy that takes account for energy resolution, and $\varepsilon(E^\prime)$ is the XENON1T detector efficiency~\cite{Aprile:2020tmw} for electron recoils.  

We show the expected energy spectrum of electrons in Fig.~\ref{sigbkg}. The orange curve shows DM-electron scattering signal spectrum at the benchmark point $m_\eta$=10 GeV, $m_\varphi$=300 GeV, $\mu$=2 keV and $\zeta=1$. The back and red curves show the background and total event spectra. The DM induced event fits in the excess above the detector energy threshold. It should be mentioned that, our model may also fit to the electron excess in the $2\mu < E_R$ scenario.
The significance of  the fittings will be given in a future study.

\section{Discussion}

We have presented  a pseudo-Dirac DM model, in which the observed relic abundance  is constituted by two component Majorana fermions with tiny mass splitting. Pseudo-Dirac DM  is stabilized by a $Z_2$ discrete flavor symmetry and  couples to the right-handed electron via a charged scalar singlet mediator.  The down scattering $\eta_b e \to \eta_a e $ with mass $\sim 10$ GeV and mass splitting $\sim 4$ keV may explain the  excess in the low energy electron recoil data recorded by the XENON1T. The issue of direct detections of inelastic DM is an interesting topic, which deserve further study in many aspects, such as the DM-electron cross section in higher order, comparison of the up-scattering case with the down scattering case, et al. These questions will be addressed in a further study.

\begin{acknowledgments}
This work was supported by the National Natural Science Foundation of China under grant No. 11775025 and the Fundamental Research Funds for the Central Universities under grant No. 2017NT17.
\end{acknowledgments}

\end{document}